\documentclass[letterpaper,twocolumn,10pt]{article}
\PassOptionsToPackage{hyphens}{url}
\usepackage{epsfig}
\usepackage{url}
\usepackage{tikz}
\usepackage{amsmath}
\usepackage{comment}
\usepackage{hyperref}
\usepackage{usenix2019_v3}

\newcommand{\FP}{first-party}
\newcommand{\FPs}{first-parties}

\newcommand{\TP}{third-party}
\newcommand{\TPs}{third-parties}
\newcommand{\TPu}{Third-party}
\newcommand{\TPus}{Third-parties}
\newcommand{\JS}{JavaScript}
\newcommand{\Web}{web}
\newcommand{\IE}{i.e.,}
\newcommand{\EG}{e.g.,}
\newcommand{\IFrame}{\texttt{<iframe>}}
\newcommand{\IFrames}{\texttt{<iframe>s}}
\newcommand{\ETAL}{\textit{et al.}}
\newcommand{\LS}{\texttt{localStorage}}
\newcommand{\IDB}{\texttt{indexDB}}
\newcommand{\EPO}{eTLD+1}
\newcommand{\subsubsubsection}[1]{\par\smallskip\noindent\textbf{#1.}}
\newcommand{\WebDataSet}{Tranco 1k}
\newcommand{\TrancoVersion}{Tranco list snapshot JZZY}
\newcommand{\NumStoragePolicies}{four}
\newcommand{\NumProfiles}{8}
\newcommand{\NumSiteURLs}{3,419}
\newcommand{\TotalNumSiteVisits}{27,352}
\newcommand{\TotalPGFiles}{280,219}
\newcommand{\TotalPGBytes}{405 GB}  
\newcommand{\PatchLinesChanges}{276}
\newcommand{\GraderAggrPerc}{95.33\%}
\newcommand{\CohensKappa}{0.69}

\newcommand{\efsVanilla}{permissive}
\newcommand{\efsVanillaCap}{Permissive}
\newcommand{\efsPrototype}{page-length}
\newcommand{\efsPrototypeCap}{Page-length}
\newcommand{\efsSplitKey}{site-keyed}
\newcommand{\efsSplitKeyCap}{Site-keyed}
\newcommand{\efsBlock}{strict \TP{} storage blocking}
\newcommand{\efsBlockCap}{Strict \TP{} storage blocking}
\newcommand{\efsBlockShort}{Third-party blocking}

\newcommand{\ToolName}{\efsPrototype{} storage}
\newcommand{\ToolNameCap}{\efsPrototypeCap{} storage}

\newcommand{\FNOne}{
  For completeness, we note that this isn't completely true, and
  that \texttt{HttpOnly} cookies cannot be accessed from \JS{}. But since
  \texttt{HttpOnly} doesn't provide protection against intentional tracking
  (since such trackers could just omit the \texttt{HttpOnly} instruction), we
  don't consider \texttt{HttpOnly} further in this work, and omit it from
  further discussion for concision.}
\newcommand{\FNTwo}{\url{https://html.spec.whatwg.org/multipage/webstorage.html\#the-localstorage-attribute}}
\newcommand{\FNThree}{\url{https://www.w3.org/TR/IndexedDB-2/}}

\begin{document}

\date{}

\title{\Large \bf There's No Trick, Its Just a Simple Trick: A Web-Compat and Privacy Improving Approach to Third-party Web Storage}

\author{
    {\rm Jordan Jueckstock}\\
    North Carolina State University\\
    jjuecks@ncsu.edu
    \and
    {\rm Peter Snyder}\\
    Brave Software\\
    pes@brave.com
    \and
    {\rm Shaown Sarker}\\
    North Carolina State University\\
    ssarker@ncsu.edu
    \and
    {\rm Alexandros Kapravelos}\\
    North Carolina State University\\
    akaprav@ncsu.edu
    \and
    {\rm Benjamin Livshits}\\
    Brave Software\\
    ben@brave.com
} 

\maketitle


\begin{abstract}
While much current \Web{} privacy research focuses on browser fingerprinting,
the boring fact is that the majority of current \TP{} \Web{} tracking is
conducted using traditional, persistent-state identifiers. One possible
explanation for the privacy community's focus on fingerprinting is that
to date browsers have faced a lose-lose dilemma when dealing with \TP{}
stateful identifiers: block state in \TP{} frames and break a significant number of
webpages, or allow state in \TP{} frames and enable pervasive tracking.
The alternative, middle-ground solutions that have been deployed all
trade privacy for compatibility, rely on manually curated lists, or depend
on the user to manage state and state-access themselves.

This work furthers privacy on the \Web{} by presenting a novel system for
managing the lifetime of \TP{} storage, ``\ToolName{}''. We compare \ToolName{}
to existing approaches for managing \TP{} state and find that \ToolName{}
has the privacy protections of the most restrictive current option (\IE{} blocking
\TP{} storage) but \Web{}-compatibility properties mostly similar to the least 
restrictive option (\IE{} allowing all \TP{} storage). This work further compares
\ToolName{} to an alternative \TP{} storage partitioning scheme inspired by 
elements of Safari's tracking protections and finds that \ToolName{} 
provides superior privacy protections with comparable \Web{}-compatibility.

We provide a dataset of the privacy and compatibility behaviors observed
when applying the compared \TP{} storage strategies on a crawl of the
\WebDataSet{} and the quantitative metrics used to demonstrate that
\ToolName{} matches or surpasses existing approaches. Finally, we provide
an open-source implementation of our \ToolName{} approach, implemented
as patches against Chromium.
\end{abstract}
\section{Introduction}
\label{sec:intro}

Web trackers use
a variety of techniques to track and violate privacy on the \Web{}. Tracking is
usually done through a mix of stateful tracking (\IE{} storing and
transmitting unique identifiers in the browser) and stateless tracking, or
fingerprinting (\IE{} attempting to uniquely identify a browser based on unique
configuration and execution environment characteristics).

Though much recent privacy work has focused on stateless, fingerprinting-based
tracking, we expect that the majority of tracking is still done using
traditional stateful methods. This intuition is based on multiple factors,
such as adtech uproar over Google's recent announcement~\cite{chromeBlogBlockingCookies}
to stop sending cookies (only one of many ways of storing identifiers) to \TPs{} in the future,
prior research demonstrating the popularity of storage-based tracking~\cite{englehardt2016openwpm, zimmeck2017privacy, papadopoulos2018cost, papadopoulos2019cookiesync, fouad2020missed, stopczynski2020redirecttracking},
and expert insight from browser developers.

While the privacy community has had some success in designing defenses
to stateless, fingerprinting tracking that protect users without breaking
benign, user-serving page functionality\cite{laperdrix2017fprandom,
nikiforakis2015privaricator}, researchers, industry and activists have been
less successful in designing practical, robust defenses against webscale
stateful \TP{} tracking.

Despite the press and attention that the ``end of \TP{} cookies'' has
received, blocking \TP{} cookies (\IE{} not sending cookies on
requests for \TP{} sub-resources) does not provide any fundamental
protections against stateful \TP{} tracking. Blocking \TP{} cookies is a positive
step for \Web{} privacy, but because it prevents categories of accidental
tracking or information disclosure, not because it prevents intentional
tracking. \TPu{} frames can access the same cookies\footnote{\FNOne{}},
\LS{}\footnote{\FNTwo{}}, \IDB{}\footnote{\FNThree{}},
or other \JS{} accessible storage methods (sometimes collectively called "DOM Storage").
In short, blocking \TP{} cookies
is a necessary, but insufficient part of solving the general problem of
preventing stateful \TP{} tracking.

Though some browser vendors have taken some steps to address \TP{} stateful
tracking, each approach has significant shortcomings and limitations.
The details of each technique are described in
Section~\ref{sec:background:browsers}, but at a high level, deployed approaches
are incomplete and insufficient, either because they depend on curated lists and
heuristics (\IE{} Firefox and Edge), address tracking across
sites but not time (\IE{} Safari), defer the question to non-expert users
(\IE{} Storage Access API\footnote{\url{https://developer.mozilla.org/en-US/docs/Web/API/Storage_Access_API}}), or provide strong protections against tracking but break sites for
users (\IE{} Brave).

We argue that practical, robust protections against stateful, \TP{} tracking
should have at least three properties.

\begin{enumerate}\itemsep=1pt
  \item \textbf{Cross-site protection}: prevent \TPs{} from using
    stored identifiers to link browsing behavior across \FP{} sites.
  \item \textbf{Cross-time protection}: prevent \TPs{} from using stored
    identifiers to link browsing behavior on the same \FP{} site across
    time.
  \item \textbf{Web Compatibility}: not effect, or minimally impact,
    user-serving, non-privacy harming behavior in \TP{} frames.
\end{enumerate}
In this work we aim to improve \Web{} privacy by presenting a new method of
managing and limiting \TP{} state that we call ``\ToolName{}''.
Section \ref{sec:design:policy} presents the approach in detail, but
at a high level, ``\ToolName{}'' is the unique combination of two features:

\begin{enumerate}\itemsep=1pt
\item 
\emph{\ToolName{} partitions \TP{} state by the top level document}.
If a browser tab has loaded a page from origin A,
and that page includes two sub-documents (\IE{} \IFrame{}s) from origin B,
the two sub-documents see the same storage, but different storage than B sees
when B is the top-level document, and also different storage than origin B
sub-documents on other pages and tabs.

\item \emph{\ToolName{} sets the lifetime of all \TPs{} state to be equal
to the lifetime of the top level document}.  If a page from origin A opens and
closes \IFrames{} from origin B, all of those origin B frames see the same
storage, even between frames being opened and closed. However, once the top-level
page is closed, all the partitioned storage for B is cleared as well.
Revisiting or reloading the top-level page will result in the B frames
seeing empty storage.
\end{enumerate}

\newcommand{\point}[1]{\par\smallskip\noindent{\bf #1.} }
\point{Contributions}
More concretely, this work makes the following contributions to improving
privacy on the \Web{}.

\begin{enumerate}
  \item The design of \ToolName{}, \textbf{a novel approach to managing \TP{} state} in
    \Web{} pages that provides strong privacy protections without breaking websites.
  \item New, general \textbf{metrics for measuring the privacy and
    web-compatibility} properties of \TP{} storage policies.
  \item An \textbf{open-source, prototype implementation} of
    \ToolName{} as a set of patches to Chromium~\cite{repo}.
  \item A public \textbf{dataset of applying \NumStoragePolicies{} storage
    policies to the \WebDataSet{}\cite{pochat2018tranco}}, a research-focused ranking of popular sites.
    Our dataset~\cite{repo} includes the  above privacy and compatibility metrics generated from  \NumStoragePolicies{} policies, each approximating a \TP{} storage policy currently deployed in a popular browser.
\end{enumerate}
\section{Background \& Motivation}
\label{sec:background}

Modern browser technologies and the security policies that govern them are complex, so we must clearly define our terms and provide essential background on browser storage policy, user tracking techniques, and the state of the art in tracking countermeasures.

\begin{figure*}[t]
    \centering
    \includegraphics[width=0.90\textwidth]{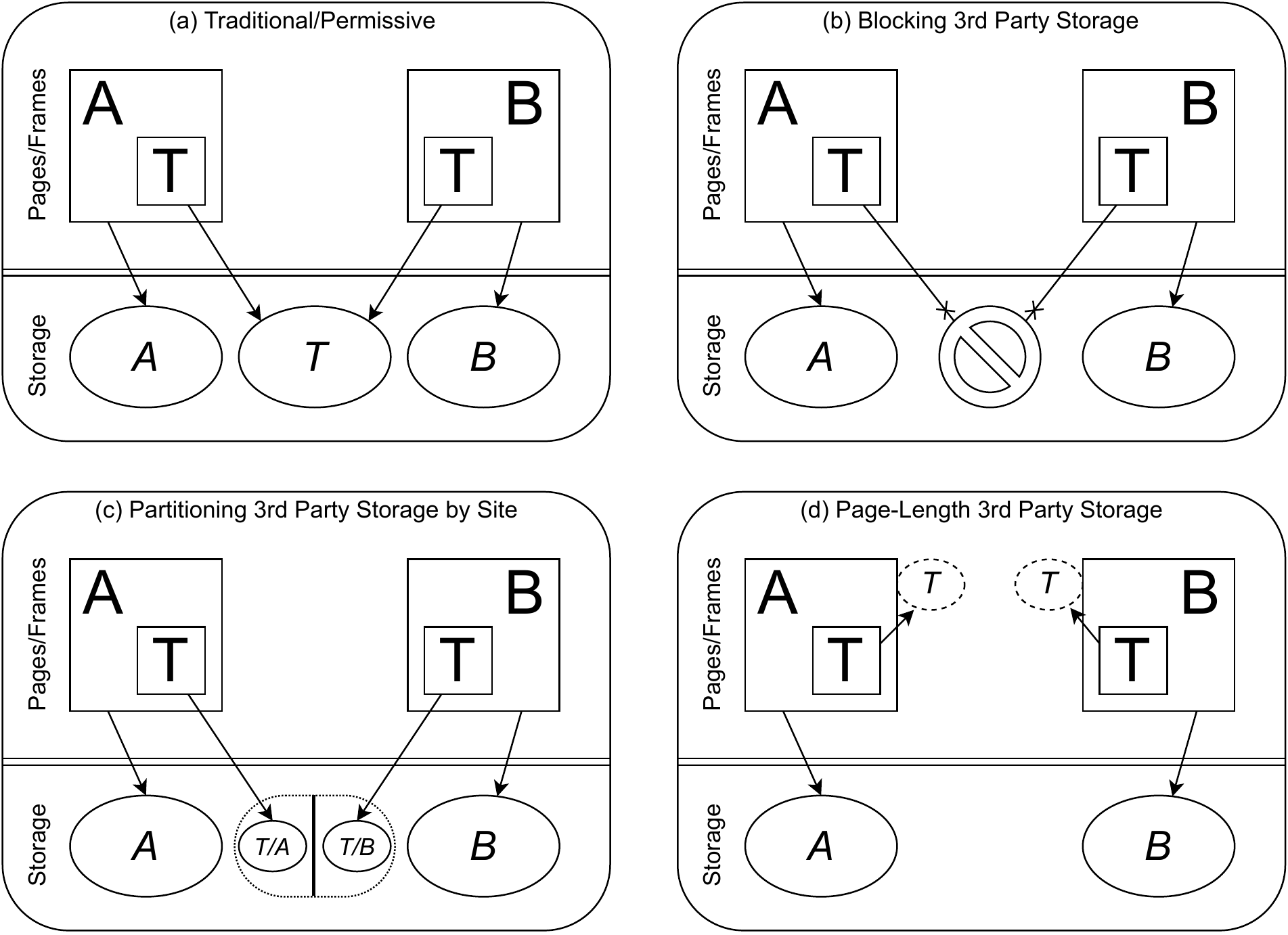}
    \caption{\TPu{} storage (a) fully allowed, (b) fully blocked, (c) partitioned by \FP{} context, and (d) scoped to hosting page life time (our proposal).  \textit{A, B, \& T are distinct domains; T is embedded as a \TP{} within A \& B.}}
    \label{fig:background:policies}
\end{figure*}

\subsection{Same-Origin Policy \& Storage Basics}
\label{sec:background:sop}

\point{Sites \& Origins}
Browsers isolate storage (\EG{} cookies, \LS{}, \IDB{}) according to the Same-Origin Policy (SOP)~\cite{whatwgSameOriginPolicy}.
The SOP has grown complex, multifaceted, and inconsistent~\cite{schwenk2017sop}, and applies to many aspects of the \Web{}; here we describe only its most basic and universal elements, particularly as they relate to storage.
An \textit{origin} comprises a scheme (\EG{} \texttt{https}), a complete DNS hostname, and an optional TCP port number.
All state-impacting activities in a browser are associated with an origin derived from some relevant URL.  For example, a script's execution origin is derived from the URL of the frame in which the script executes, and an HTTP request origin is derived from the URL being fetched.

Many activities are restricted to \textit{same-origin} boundaries.  For example, a script executing in origin A cannot access cookies stored for origin B.  This is true even when a sub-document from origin B is embedded in a page from origin A.
Storage is strictly isolated according to SOP: scripts can access cookies and DOM storage (\EG{} \LS{}) only for their execution origin, and HTTP requests store and transmit cookies only for their destination origin.


\point{First and Third Parties}
We now define two terms used through the rest of this paper, \FP{} and \TP{}.  These terms are not unique to this work, but are frequently used to mean similar but not-quite-the-same things in research and \Web{} standards, so we define their use in this work explicitly.

When loading a website, the \textbf{\FP{}} is the ``site'' portion of the top level document.  This is the \EPO{} of the URL shown in the navigation bar of the browser.  Any sub-resources or sub-documents included in the page are considered \FP{} if they're fetched from the same \EPO{} as the top level document. \textbf{Third-parties}, then, are any site not equal to the top-level document. A sub-document (\IE{} \IFrame{}) is considered \TP{} if it is fetched from any origin not-equal to the top-level document, a script is \TP{} if it was fetched from a site different than the top level document, and so forth.

Finally, we note that when applying SOP to the \Web{}, what determines storage access is the ``site'' of the frame including a script, not the script itself.  So if a page from origin A includes a script loaded from origin B, the script is \TP{}, but has access to the \FP{} storage. What storage area a script has access to is determined by the ``site'' of the page, not the ``site'' of the script.

\subsection{User Tracking}
\label{sec:background:tracking}

\point{Types of Behaviors Tracked}
This work uses the term ``tracking'' to refer to a \TP{} re-identifying a visitor across visits to \FP{} sites. Unless otherwise specified, we use ``tracking'' to refer both to \textbf{cross-site tracking} (\IE{} a \TP{} can link an individual's behavior across \FPs{}) and \textbf{cross-time tracking} (\IE{} a \TP{} can identify the same person returning to the same \FP{} across sessions).


\point{Stateful Tracking}
The oldest, simplest and most common form of online tracking is ``stateful'' tracking, where a \TP{} stores a unique value on the user's browser, and reads that value back across different \FPs{}.  While the terms \textit{explicit} and \textit{inferred} used by Roesner \ETAL{}~\cite{roesner2012detecting} appear more precise, as both techniques involve state of some kind, the \textit{stateful}/\textit{stateless} terminology popularized by Mayer and Mitchell~\cite{mayer2012fourthparty} appears dominant in subsequent research.

In the simplest case, stateful tracking works as follows. Sites A and B both include an iframe embedding site C.  When the embedded site C is loaded, it looks to see if a unique identifier has been set, and if not it generates and stores one, using any of the storage methods provided by the browser (cookies, \LS{}, \IDB{}, etc). Embedded site C then checks to see what site's its being embedded in, and sends a message back to site C, recording that the same user visited both sites A and B.

Stateful tracking, at root, relies on a site being able to access persistent state in different contexts, and using the persistently stored state to link (conceptually) unrelated behavior. To build on the previous example, site C is able to track the user across A and B because C seems the same storage values when embedded in sites A and B, even though site A might not have any direct relationship with site B.

As we will discuss at length later, approaches for preventing stateful tracking involve either preventing \TPs{} from storing values, proving \TP{} different stored values when embedded in different contexts, or combinations of the two.

\point{Other Tracking Techniques}
While stateful tracking is the simplest, and likely the most common, method for tracking users online, there are other ways \TPs{} track users. While this work focuses on stateful tracking, in this subsection we briefly discuss these other, non-stateful techniques here for completeness:
\begin{itemize}\itemsep=-1pt
\item \emph{Browser fingerprinting} refers to uniquely identifying a browser (or browser user) not through the storage and transmission of a unique identifier, but by identifying unique characteristics of the browser's configuration (\EG{} plugins, preferred language, ``dark mode'') and execution environment (\EG{} operating system, hardware capabilities).

\item \emph{Server-side tracking} is a broad term that loosely means tracking users across sites not through stored identifiers (\IE{} stateful tracking) or unique configuration (\IE{} fingerprinting), but through information the user provides to the site.  For example if a user uses the same email address when registering on two different sites, a tracker could later use the repeated email address to link the users behavior across sites.
\end{itemize}

\subsubsection{Focus on Stateful Tracking}
This work presents a novel solution for preventing stateful cross-site and cross-time tracking.  We aim to improve protections against stateful tracking because we think its where browsers are most lacking practical, robust, comptable defenses. While significant research has gone into building \Web{}-compatible defenses against stateless tracking (\EG{} \cite{laperdrix2017fprandom, nikiforakis2015privaricator}), the existing techniques for preventing stateful \TP{} tracking are either incomplete (\IE{} they still allow significant privacy harm to occur) or incompatible (\IE{} they break a significant number of websites).

\subsection{Deployed Stateful Tracking Defenses}
\label{sec:background:browsers}

Real-world countermeasures currently deployed in production browsers illustrate a range of possible tradeoffs between privacy and compatibility.
We rely heavily on the community-curated Cookie Status project~\cite{cookieStatusProject} for up to date policy implementation details.

\point{Brave: Block all Third-Party State}
Brave~\cite{braveBrowserWebsite}, a Chromium fork featuring aggressive privacy protections called ``Shields'', defaults to blocking all forms of \TP{} storage.
The officially correct way to block persistent \TP{} storage involves raising a \JS{} exception on script access to blocked storage APIs~\cite{whatwgStorageBlocking}.
Few sites are prepared to handle these exceptions, however, and Brave improves compatibility by instead simply turning blocked \TP{} storage API accesses into no-ops.
Brave also uses a whitelist to allow a small number of high-profile \TPs{} to use persistent storage in the context of specific \FP{} sites (\EG{} \textit{googleusercontent.com} when embedded from \textit{google.com})~\cite{githubBraveWhitelist}.
Brave's approach results in strong privacy protections at the cost of a higher incidence of site breakage, which may require users to selectively lower its Shields on incompatible sites.

\point{Safari: Partition Third-Party State}
Apple Safari~\cite{safariBrowserWebsite} features Intelligent Tracking Prevention (ITP), a combination of storage restriction policies, opt-in APIs, and on-client classification of tracking domains via machine-learning~\cite{webkitITPintro}.
Safari never transmits cookies on \TP{} HTTP requests.
Cookie and \texttt{localStorage} access in \TP{} frames are partitioned on \FP{} site identity to prevent stateful lateral tracking (as in Figure~\ref{fig:background:policies}c).
Safari provides developers with a \texttt{requestStorageAccess} API to request user permission to access unpartitioned \TP{} storage across \FP{} contexts.
This opt-in approach allows users to accept the potential for lateral tracking in exchange for useful functionality such as cross-site login state.

Safari features a number of additional policies and heuristics to restrict the lifetime of items stored by domains ITP has classified as probable trackers.
These restrictions impact but do not categorically eliminate potential for longitudinal tracking by \TPs{}.
The Safari ITP approach provides strong cross-site tracking protections while avoiding full-blocking with its associated site breakage, but it does not eliminate across-time tracking by \TPs{}, and the non-deterministic impact of machine-learning on its policy enforcement can make it challenging for web developers to reason about.

\point{Firefox and Edge: Restrict Known Bad Actors}
Mozilla Firefox~\cite{firefoxBrowserWebsite} has adopted a selective storage policy that depends on the Disconnect~\cite{disconnectTrackerList} list of curated tracking domains.
In general, \TP{} origins not found in the Disconnect list are granted unrestricted access \TP{} storage.
\TPu{} origins classified as trackers by Disconnect are given access to \TP{} storage on the first five \FP{} sites embedding that \TP{} origin.
Subsequent additional sites embedding that \TP{} origin will result in user opt-in prompts to allow \TP{} storage which must be accepted to allow use of \TP{} storage by that origin on that \FP{} site.

Exceptions to these restrictions are made for \FP{} domains identified by the Disconnect list as related to specific \TP{} origins (\EG{} \textit{googleusercontent.com} embedded on \textit{google.com}).
The Firefox approach is one of compromise: well-known tracking domains face restrictions on the reach of their lateral tracking, but protection depends heavily on the validity and coverage of the underlying filter list.

Microsoft Edge~\cite{edgeBrowserWebsite} has begun to deploy filter list-based storage restrictions similar to those performed by Firefox, with all the benefits and drawbacks of this compromise approach summarized above.

\point{Chrome: Unrestricted Third-Party State}
Google Chrome~\cite{chromeBrowserWebsite}, in contrast to all of the above, permits full \TP{} storage use, including sending cookies on HTTP requests to \TP{} resources.
Google has announced intentions to phase out \TP{} cookie support~\cite{chromeBlogBlockingCookies} in the near future; technical details remain vague, but their wording implies eliminating only cookies on \TP{} HTTP requests, not restricting \TP{} storage in general.
Chrome dominates as the world's most popular browser for both desktop and mobile markets~\cite{statCounterBrowserPopularity}, understandably prompting \Web{} developers to target its behavior for maximum compatibility, and indirectly perpetuating the status quo of stateful tracking in the process.

\subsection{Compatibility and Tracking Protections}
Finally, we present some ways that existing protections against \TP{} stateful tracking break websites.  We present these as moderating examples, and useful constraints, in designing \ToolName{}. Without considering these compatibility concerns, solutions will tend to simplistic, ``block-everything'' approaches that end up not being useful, and so not being effective in protecting privacy.

We gather the following examples from Brave's public issue tracker\footnote{\url{https://github.com/brave/brave-browser/issues}}. We pull from Brave's breakage reports since Brave has the most aggressive restrictions on \TP{} storage of the surveyed browsers, and so the largest number of storage-related compatibility problems.  Nevertheless, we present these as examples of the kinds of compatibility problems that \TP{} storage protections can cause.

\subsubsection{Uncaught Exceptions from Blocking Storage}
\efsBlockCap{} breaks embedded SlideShare slide show widgets (\EG{} \url{https://support.blogactiv.eu/2015/04/24/how-to-embed-slideshare/}) on Chrome.
The widget becomes inert, not responding to clicks, when Chrome's implementation of \efsBlock{} (correctly, per the specification) raises \JS{} exceptions on access to storage APIs.
Brave's silent no-op implementation of \efsBlock{} is sufficient to prevent breakage in this case; successful storage access is clearly not essential to this widget's functionality.

\begin{figure}
    \centering
    \includegraphics[width=\columnwidth]{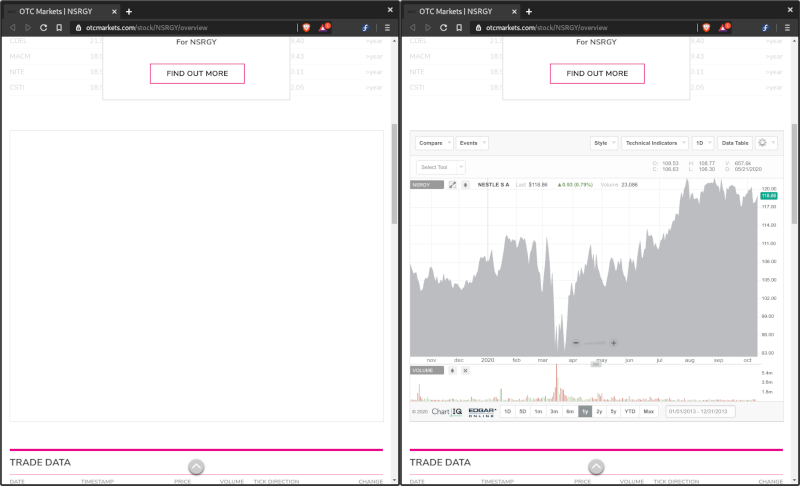}
    \caption{Stock market graph broken by \efsBlock{} (left) and working with \ToolName{} (right).}
    \label{fig:discussion:broken-graph-side-by-side}
\end{figure}

A similar example is provided by a data plot widget broken by \efsBlock{} (\EG{} \url{https://www.otcmarkets.com/stock/NSRGY/overview}).
Once again, \efsBlock{} causes a \JS{} run-time error which results in a blank data plot (see Figure~\ref{fig:discussion:broken-graph-side-by-side}).
In this case, Brave's silent no-op blocking does not help: the error is caused by property access on a null value returned from a no-op API stub.

\subsubsection{Breaking Cookie-Based Third-Party Sessions}
A server-side example of \efsBlock{} causing breakage is provided by a live code editing/running widget embedded in the R language documentation (\EG{} \url{https://www.rdocumentation.org/packages/grid/versions/3.6.2/topics/grid.plot.and.legend}).
The embedded widget tries to establish a cookie-based session with \TP{} domain \textit{multiplexer-prod.datacamp.com}.
Failure to persist \TP{} cookies results in HTTP 403 errors on subsequent HTTP requests, preventing code execution and output display.

A broken video player on a popular commentary and analysis site provides another example (\url{https://fivethirtyeight.com/videos/do-you-buy-that-biden-should-pick-a-running-mate-from-a-swing-state/}).
With \TP{} storage blocked, the video player remains blank indefinitely. In this case, the video player functionality is broken because the frame attempts to use \LS{} to persist values across pages.

\section{Design \& Implementation}
\label{sec:design}

We propose and prototype a novel browser storage policy that prevents both cross-site and cross-time stateful tracking while measurably improving site compatibility over traditional \TP{} storage blocking.
\ToolNameCap{} prevents stateful \TP{} tracking while minimizing site breakage by making \TP{} storage fully functional within a strictly isolated, ephemeral scope.
We also provide a brief overview of how we developed and tested our \ToolName{} prototype within a Chromium-based browser.

\subsection{Policy Design}
\label{sec:design:policy}

The key insight behind \ToolName{} is that site breakage can be minimized without compromising privacy by making all interactions with \TP{} storage behave normally (\IE{} permissively), but only within the isolated, ephemeral scope of the containing page's lifespan.
The containing page is the top-level frame, loaded from the URL displayed in the browser's navigation UI (\EG{} address bar).
Its lifespan expands from the moment the top-level frame committed to loading that document URL to the moment any navigation event (including even reloads of the same URL) discards the contents of the top-level frame.

Within the isolated, ephemeral scope of each top-level page object's lifespan, \TP{} storage access behaves normally for both scripts executing in \TP{} \IFrames{} and HTTP requests to \TP{} domains.
SOP enforcement is unchanged.
All storage behaves in the traditional, permissive way with one exception: \TP{} storage starts out empty and is discarded along with the top-level page object on top-frame navigation.
Any site embedding \TP{} content which functions correctly under \efsVanilla{} policy for first-time visitors with empty cookie jars should function correctly under \ToolName{} policy.

Isolating \TP{} storage to single page lifespans provides a good tracking \textit{vs.} compatibility compromise.
\ToolNameCap{} prevents stateful cross-site and cross-time tracking automatically, as \TPs{} cannot ``remember'' anything past top-level page (re-)loads.
Compare Figures~\ref{fig:background:policies}a and \ref{fig:background:policies}d.
As a practical matter, \TP{} content cannot silently manipulate top-level page navigation, so it cannot test whether \TP{} storage will persist beyond top-level navigations.
All tests that can be done silently within the scope of a single page's lifespan will appear fully functional, as for a first-time visitor with uninitialized \TP{} storage.

Some hypothetical examples illustrate the impact of this approach.

First, consider two \IFrames{} from the same \TP{} embedded on a single page document: these will share the same ephemeral \TP{} storage partition and so can use all forms of \TP{} storage, via both script access and HTTP cookies, to communicate with each other and the remote \TP{} origin for the duration of the embedding page's lifespan.

Second, consider a \TP{} \IFrame{} embedded on a page that is loaded on two different tabs simultaneously: each instance of the frame is using its parent-page's ephemeral \TP{} storage partition, so no cross-site stateful sharing/communicating is possible between them.

Third, consider a \TP{} \IFrame{} embedded on a page that is loaded and then reloaded in the same tab: each page load (regardless of URL) discards the previous page's ephemeral \TP{} storage partition, so no cross-time stateful sharing/communicating is possible.

Finally, consider a \TP{} \IFrame{} embedded in two pages hosted on different \FP{} domains: whether these pages are loaded sequentially in one tab, or simultaneously in two tabs, each \TP{} frame is using its own parent-page's ephemeral \TP{} storage partition, so again no cross-site stateful sharing/communicating is possible.

\subsection{Prototype Implementation}
\label{sec:design:prototype}

We implement our \ToolName{} prototype as a set of patches to Brave 1.12.48 (based on Chromium 83).
We use Brave, and version 1.12.48 specifically, in order to use the latest revision of PageGraph for data collection, per Section~\ref{sec:methodology:collection:pagegraph}.
However, our patches are completely independent of PageGraph's patches and can be built without them present.
The most relevant change from stock Chromium provided by Brave is its gentler approach to \TP{} storage blocking, which it enables by default.
Instead of raising a \JS{} exception on script access to blocked storage (per the specification), Brave makes the access a silent no-op, returning a null value.

Chromium's standard architecture includes content and storage isolation mechanisms relevant to our design goals.
In addition to classic SOP enforcement, Chrome isolates content rendering and \JS{} execution into separate render processes partitioned on a same-site basis (see Section~\ref{sec:background:sop}).
Each render process uses \textbf{one} storage partition, which can be persistent (the default) or ephemeral (private mode), and which can additionally be partitioned by arbitrary identifiers (for Chrome apps and extensions).
HTTP traffic is managed by a dedicating networking process, which chooses a storage partition for HTTP cookies based on the frame initiating the request.

We exploit this existing site storage isolation framework to prototype \ToolName{} with minimal changes to the browser.
Our classification of frames and requests as \TP{} reuses the same-site logic already in Chromium and is always relative to the top-level page URL (not \IFrame{} URLs).
Each time a tab's top-level frame loads a page URL, we generate and store a UUID identifying that load event (the \textit{load key}).
When \TP{} frames are subsequently created and assigned to separate render processes, we augment the \TP{} site identifier with the top-level frame's current load key to enforce page-level isolation between ephemeral \TP{} storage partitions.
HTTP requests to \TPs{} are bound to the associated ephemeral \TP{} storage partition, which is created on demand if necessary.
The result is unchanged \FP{} storage behavior, and fully functional \TP{} storage that lives only as long as the containing page document, as in Figure~\ref{fig:background:policies}d.

\subsection{Implementation Remarks}
\label{sec:design:results}

The changes required to prototype \ToolName{} proved deceptively small.
A total of \PatchLinesChanges{} lines of C++ were added, changed, or removed by our patches.
The scale of these patches, small relative to the millions of source code lines of Chromium, belie the challenge of finding the right places to patch.
Most changes relating to storage partition creation and isolation were confined to the main ``browser'' process in Chromium's multi-process architecture, which has access to the entire frame tree for each tab, and were thus relatively straightforward to implement.
Binding \TP{} HTTP requests to isolated, ephemeral storage partitions on demand, however, crossed process boundaries into the network process, which does not have access to frame tree context information, and required additional IPC messaging and timing concerns.

The implementation demonstrated correctness and robustness fully sufficient for prototype testing.
The available Chromium unit tests all passed, except for a few implementation-specific assertions we expected to fail after our changes.
Manual testing of multiple scenarios like the examples from Section~\ref{sec:design:policy}, included nested \TP{} \IFrames{}, showed expected behavior in all cases.
Furthermore, the prototype's error rate during automated crawls were favorably comparable to stock \efsVanilla{} policy (see Section~\ref{sec:results:stats}).

Prototype performance proved adequate despite not being a design priority.
Because performance was not a design priority for the prototype, we did not perform any benchmarks.
In theory, our approach should reduce performance over stock Chromium: it can produce more render processes, can involve more I/O operations creating temporary directories, and definitely invokes additional IPC overhead between network and render processes.
However, both manual testing and automated crawling with the prototype revealed no obvious performance degradation.
Furthermore, none of these issues are inherent in the policy itself, and there is no reason to believe a performance-tuned production implementation would produce any significant overhead compared to traditional policies.

\section{Methodology}
\label{sec:methodology}

We evaluate our proposed policy by comparing its tracking and compatibility performance against alternative policies during automated, stateful crawls of popular \Web{} sites.

\subsection{Crawl Methodology}
\label{sec:methodology:collection}

Here we describe our data collection procedure in sufficient detail to permit straightforward experiment reproduction.

\subsubsection{Target URLs}
\label{sec:methodology:collection:urls}

We generated a seed list of URLs to visit in parallel using a stateless \textit{pilot crawl} of the \WebDataSet{} sites~\cite{pochat2018tranco}.
To achieve depth and representative sampling of \Web{} content, we must explore more than just the ``landing page'' of each site.
But each of our \NumProfiles{} parallel crawls must visit the same sequence of page URLs to produce comparable results.
Coordinating the link spidering and selection process across parallel crawls introduces needless engineering complexity.
Our solution was to perform a stateless pilot crawl using stock Brave to visit the \WebDataSet{} sites' landing pages and spider three links deep into the site structure.
This approach, using \TrancoVersion{}, produced \NumSiteURLs{} total deduplicated page URLs to visit.

\subsubsection{Policy Variants}
\label{sec:methodology:collection:policies}
   
We collect data using \NumStoragePolicies{} distinct policy variants:
\begin{itemize}
    \item \textbf{\efsVanillaCap{}:} Allows all forms of \TP{} storage, as per Figure~\ref{fig:background:policies}a.  Stock Chrome behavior.  Presumed to cause no breakage.
    \item \textbf{\efsBlockCap{}:} Blocks all forms of \TP{} storage, as per Figure~\ref{fig:background:policies}b.  Treats access as no-op.  Presumed to cause the most breakage.
    \item \textbf{\efsSplitKeyCap{}:} Partitions persistent \TP{} storage by \FP{} \EPO{}, as per Figure~\ref{fig:background:policies}c.  Alternative to our proposed policy, inspired by elements of Safari ITP.
    \item \textbf{\efsPrototypeCap{}:} Isolates \TP{} storage in ephemeral, per-page partitions, as per Figure~\ref{fig:background:policies}d.  Our proposed policy.
\end{itemize}

\subsubsection{Crawl Execution}
\label{sec:methodology:collection:execution}

We executed our stateful crawls in parallel across all storage policies without any simulated user interactions.
We deployed two instances of each tested policy to verify behavioral consistency and provide similarity-score baselines (see Section~\ref{sec:methodology:evaluation:compat-quantitative}).
The crawlers maintained independent, persistent user profiles for each policy instance to maintain realistic state across all sequential page visits.
The full experiment included 2 iterations crawling the master URL list to provide data on cross-time tracking across repeat visits.
All crawls were performed in parallel and simultaneously (but without active synchronization between profiles) from a single network vantage point.
Each page visit was performed in a freshly launched, non-headless (\IE{} rendering to the \textit{Xvfb} headless display server) browser instance.
Navigation was allowed to time out after 30 seconds.
Assuming no navigation timeout, our crawlers waited for 30 seconds after the \texttt{DOMcontentloaded} event (\IE{} main document fetched and parsed but subresources not fully loaded yet) before tearing down the browser instance.
No simulated user interactions were attempted.

\subsubsection{PageGraph Instrumentation}
\label{sec:methodology:collection:pagegraph}
    
We use PageGraph, an instrumentation system built into an experimental branch of Brave, to record internal page behaviors.
PageGraph patches the V8 JS engine and the Blink HTML rendering engine to capture and annotate a graph of each HTML document's DOM structure and the events that constructed and modified it.
Nodes represent entities such as DOM elements, scripts, HTTP resources, storage mechanisms, and a selective subset of builtin and DOM-provided \JS{} APIs.
Edges represent relationships between nodes such as DOM structures and script interactions with DOM elements, DOM events, \JS{} APIs, and HTTP requests.
The set of non-structural edges in each of these graphs constitute the dynamic behaviors of the originating page.
Behavioral-edge-set similarity can be quantified using Jaccard index scores to provide a useful proxy for behavioral compatibility among compared storage policies.

\subsection{Evaluation Methodology}
\label{sec:methodology:evaluation}

We evaluate our proposed policy's privacy and compatibility performance using full-scale quantitative stateful tracking metrics, full-scale quantitative site behavior similarity metrics, and randomly-sampled qualitative assessment of site breakage.

\subsubsection{Preliminary Data Filtering}
\label{sec:methodology:evaluation:prelim}

We focus our analysis on frames of interest: \IE{} \TP{} frames not flagged as advertisements.
Our classification of \TP{} \textit{vs.} \FP{} frames is based on \EPO{} matches derived from the Public Suffix List~\cite{publicSuffixListSnapshot}.
Frames loaded from the same \EPO{} as the main page URL are \FP{} frames; all others are \TP{} frames.
We eliminate from consideration all \FP{} frames and \TP{} frames flagged as advertising content by the community-maintained EasyList~\cite{easylistSnapshot}.
This filtering eliminates noise from our evaluation: \FP{} storage is not affected by our policy change, and we are unconcerned about breakage of known advertising content.

\subsubsection{Quantitative Privacy Assessments}
\label{sec:methodology:evaluation:privacy}

\subsubsubsection{Tracking Potential}
The central metric we use to quantify potential for stateful cross-site and cross-time tracking by \TPs{} is the \textit{potentially identifying cookie flow} (PICF).
A \textit{cookie flow} is the combination of an HTTP cookie and a \TP{} \EPO{} receiving that cookie.
We consider cookie flows \textit{potentially identifying} when the cookie values meet a tunable minimum size threshold and are \textbf{unique} to a single browser profile during our stateful crawls.
There are other forms of \TP{} storage available, and other channels by which identifying tokens can be transmitted to \TPs{}.
But we use cookies as a representative measure of stateful tracking because they are unambiguous in structure, ubiquitous as tracking IDs, and essentially unrestricted by stock Chrome, our baseline.
(Both our \ToolName{} and \efsSplitKey{} implementations apply their storage policies to \textbf{all} forms of \TP{} storage, not just cookies.)

\subsubsubsection{Cross-Site Tracking}
Identical PICFs seen across multiple distinct top-level sites visited represent potential for cross-site tracking by the associated \TP{} domain.
We aggregate cross-site PICFs to count the total number of top-level sites across which each distinct \TP{} domain seen could have tracked our crawler profiles, giving us summary scores of ``cross-site trackability'' by which to compare all our storage policies.
These scores can be visualized using cumulative sum curves, as shown in Section~\ref{sec:results:privacy:lateral}.

\subsubsubsection{Cross-Time Tracking}
PICFs seen on a given top-level site across multiple pages/crawls represent potential for cross-time, or visit-to-visit, tracking by a given \TP{} domain.
We aggregate cross-time PICFs to count the total number of \TP{} domains which could have tracked our crawler profiles for each distinct top-level site domain visited, giving us summary scores of ``cross-time trackability'' by which to compare all our storage policies.
These scores can be visualized using cumulative sum curves, as shown in Section~\ref{sec:results:privacy:longitudinal}.

\subsubsection{Quantitative Compatibility Assessment}
\label{sec:methodology:evaluation:compat-quantitative}

We assess site compatibility across storage policies using a quantifiable proxy measure: similarity of internal page behaviors as reported by PageGraph.
Our insight is to presume no storage-based breakage for \efsVanilla{} profiles and some unknown (but non-zero) amount of breakage on \efsBlock{} profiles.
If alternative policy (\EG{} \ToolName{}) profiles produce content behaviors more similar to the \efsVanilla{} baseline than do the \efsBlock{} profiles, then the alternate policy is less likely than \efsBlock{} to cause breakage.

We model and compare content behaviors using the set of non-structural (i.e., action or event) edges in PageGraph representations of relevant frames.
Similarity between edge sets can be measured using the Jaccard index: $J(A, B) = \frac{|A \cap B|}{|A \cup B|}$.
Index scores range from 0 (no intersection) to 1 (equality).
We consider the score undefined when both sets were empty.

We compare content behaviors across identical frames loaded on identical pages across all tested policies.
Frames and pages are identified and matched by full URL.
The similarity score of the two \efsVanilla{} profiles provides the compatibility baseline: the presumed best-possible similarity score for that frame/page instance.
The other profiles are each compared with a single \efsVanilla{} profile to provide similarity scores to compare against the baseline.
The cumulative sum of all frame/page instance similarity scores for each profile can be visualized to show which policies track closest to the baseline across all visited pages (see Section~\ref{sec:results:compat:quantitative}).

We optimized the set of PageGraph node types included in our behavioral sets to maximize the distance between \efsBlock{} policy scores and the \efsVanilla{} baseline score.
Our intuition is that the baseline score provides a threshold of ``reasonable'' behavioral differences between two different instances of the same content loaded in different browsers at about the same time.
The farther away from this baseline a policy scores, the greater the likelihood of unreasonable, or breaking, differences in behavior.

We identified 11 PageGraph node types relevant to behavioral analysis, a set small enough to be amenable to brute force optimization across its power set.
Optimization relied on a random sample of 100 frame/site instances extracted from a preliminary full-scale crawl dataset, whose unoptimized similarity curves matched those of the entire data set, indicating a representative sampling.
On this data subset we tested the \efsBlock{} separation from the \efsVanilla{} baseline for every subset of relevant PageGraph node types.
The results confirmed our intuition that the least helpful node types were structural elements like HTML elements and DOM text blocks; less intuitively, they also showed that PageGraph's set of instrumented DOM manipulation \JS{} APIs was similarly unhelpful.
The final optimial node type set comprised scripts and PageGraph's selected \JS{} builtin APIs (\EG{} date functions), HTTP resources, frame structures (DOM roots and frame-owning elements), and storage mechanisms (cookie jars, local and session storage buckets).
Only edges (\IE{} behaviors) linking these node types are included in the behavior similarity results presented in Section~\ref{sec:results:compat:quantitative}.

\subsubsection{Qualitative Compatibility Assessment}
\label{sec:methodology:evaluation:compat-qualitative}

We augment our quantitative proxy assessment of site compatibility with blinded multi-grader manual analysis for breakage within a random sample of sites loading popular \TP{} content.
Our methodology is heavily inspired by a similar experiment by Snyder \ETAL{}\cite{snyder2017most}.

To select our set of URLs to test, we first identified the most popular \TP{}, non-ad-blocked frame URLs within our crawl dataset. We sorted these by the harmonic mean of the number of pages embedding that frame and the number of \TP{} cookies set for the frame's \EPO{}. This metric is higher for frames which appear on a large number of sites and have access to a large number of cookies: prime candidates for testing \TP{} storage policy changes. We selected the top 10 frame URLs with distinct \EPO{}s, to have higher content diversity. We further filtered out frames which appeared only on non-English sites (\EG{} frames from \textit{baidu.com} and \textit{alicdn.com}), and frames which did not have a content type of either HTML or \JS{} (\EG{} frames from \textit{sharethis.com} with a content type of image).

For each of the 10 selected frame URLs, we randomly selected 5 candidate page URLs observed to embed that frame URL during our crawls, giving us 50 candidate URLs for manual analysis. Upon closer inspection of the frame contents, some frames did not have any real estate on the page and simply contained JS script, which would interact or render with DOM elements elsewhere on the containing site. With this insight, we adopted a holistic approach to evaluate breakage rather than simply observing the behavior of one frame.

We had two graders evaluate each of our candidate URLs for the policy variants in Section~\ref{sec:methodology:collection:policies}. The graders would visit a candidate URL first with a \efsVanilla{} profile, the Chrome default. This visit is our \textit{control} visit for manual analysis. It was followed by a visit to the same URL with each of the \efsSplitKey{}, \efsPrototype{}, and \efsBlock{} profiles. Every visit, including the control visit, was independent of all others, with a fresh browser profile to ensure no browsing state carried over between tests/visits. To keep our graders unbiased, subsequent visits to the candidate URL after the control visit were randomly coded so the graders did not know which profile they were using.

In our holistic approach to evaluation, each grader would visit the candidate URL with the control profile first. We instructed each grader to perform as many interactive actions on the candidate site within one minute, which is the average dwelling time for a typical web-user on a webpage~\cite{liu2010understanding}. Activities depended on category of site: on news portals, our graders would skim through, search for articles, watch embedded videos, click on ads, or try to sign-up for newsletters; on shopping sites, they would search for products, add products to the shopping cart, and initiate a checkout; on product sites, graders would either skim informational material, or try any video streams available, etc. Subsequently, the graders would visit the same URL with the 3 coded profiles, performing similar actions as during the control visit, observing any deviations from the control visit, and scoring their visit on a 1 to 3 scoring scale. The graders gave a coded profile visit a score of \textbf{1} if the visit did not have any perceptible deviations from the control; \textbf{2} if there were some deviations from the control visit, but this did not hinder their visiting experience or the tasks the graders attempted on the site; and \textbf{3} if the visit had significant deviations from the control, preventing the graders from replicating their control visit activities.

Given the highly subjective nature of the evaluation scheme, we carefully assess grader agreement.
Our graders evaluated the candidate URLs independently, unaware of the other grader's scores.
In our evaluation, our graders had a high agreement percentage (~\GraderAggrPerc{}).
We also computed the Cohen's Kappa inter-rater reliability statistic~\cite{hallgren2012computing} as~\CohensKappa{}, showing statistically substantial agreement between our graders~\cite{mchugh2012interrater}.
We present the results of our manual evaluation in Section~\ref{sec:results:compat:qualitative}.

\section{Results}
\label{sec:results}

Our experimental results show that \ToolName{} combines best-case stateful tracking protection with near-best-case site compatibility.

\subsection{Crawl Statistics}
\label{sec:results:stats}

Our stateful \Web{} crawls ran from September 12-16 on a single Linux virtual machine (40 VCPUs, 100GiB RAM).
Combined, the crawls visited \TotalNumSiteVisits{} total pages using \NumProfiles{} user profiles and produced \TotalPGFiles{} PageGraph files (\TotalPGBytes{}).

\begin{figure}[t]
    \centering
    \includegraphics[width=\columnwidth]{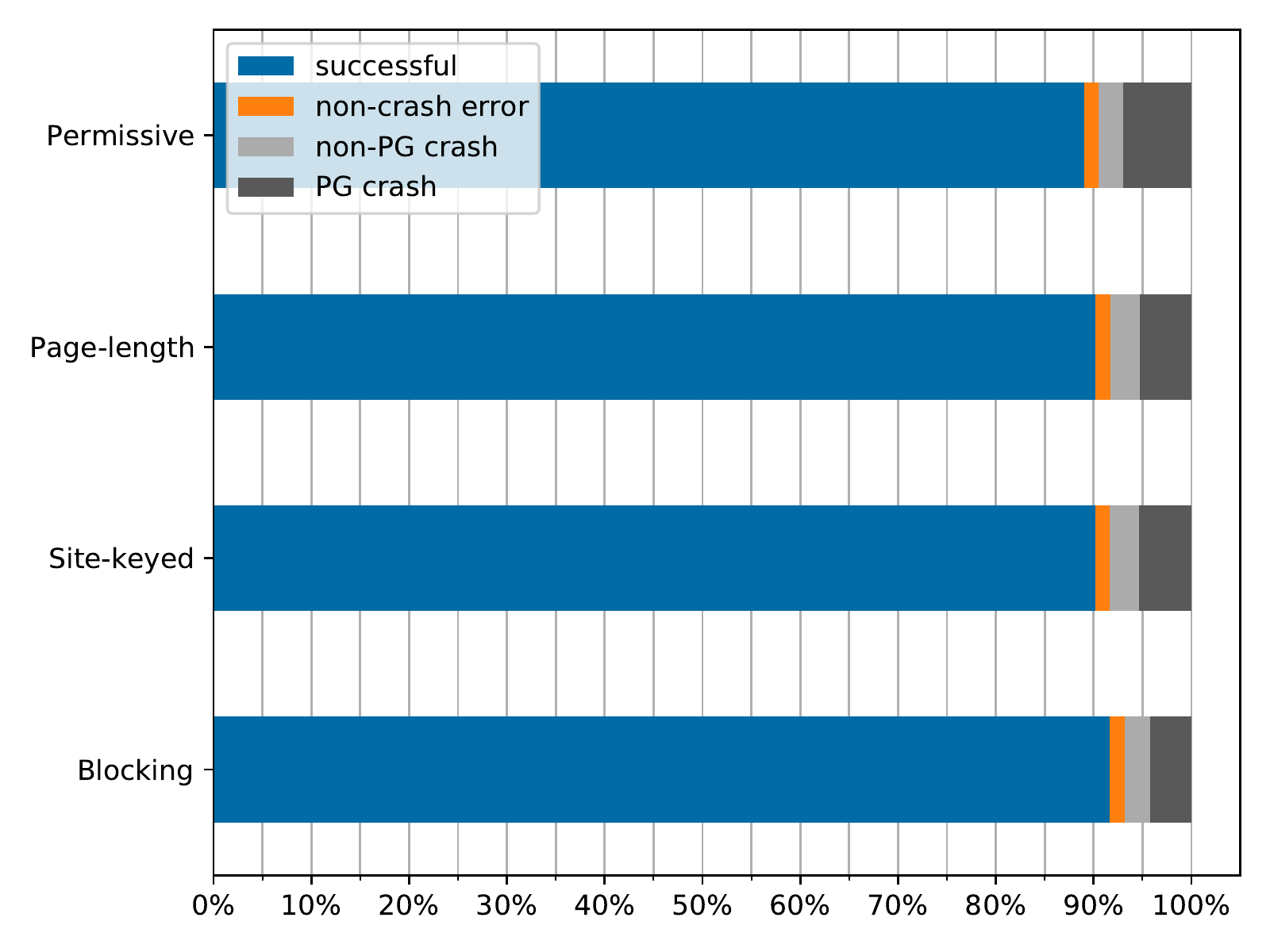}
    \caption{Crawl success rate varied modestly across policies but was always reasonably high.}
    \label{fig:results:errors}
\end{figure}

Error rates were acceptable (Figure~\ref{fig:results:errors}) if somewhat amplified by PageGraph internal consistency assertion failures.
PageGraph's instrumentation is expansive and tracks complex interactions between \JS{} execution, DOM manipulation, and network traffic.
Whenever unexpected corner cases (or bugs) prevent it from establishing unambiguous context for an event or activity, PageGraph logs the issue and terminates the browser rather than recording unreliable data.

\subsection{Privacy: Cross-Site Tracking Potential}
\label{sec:results:privacy:lateral}

\begin{figure}[t]
    \centering
    \includegraphics[width=\columnwidth]{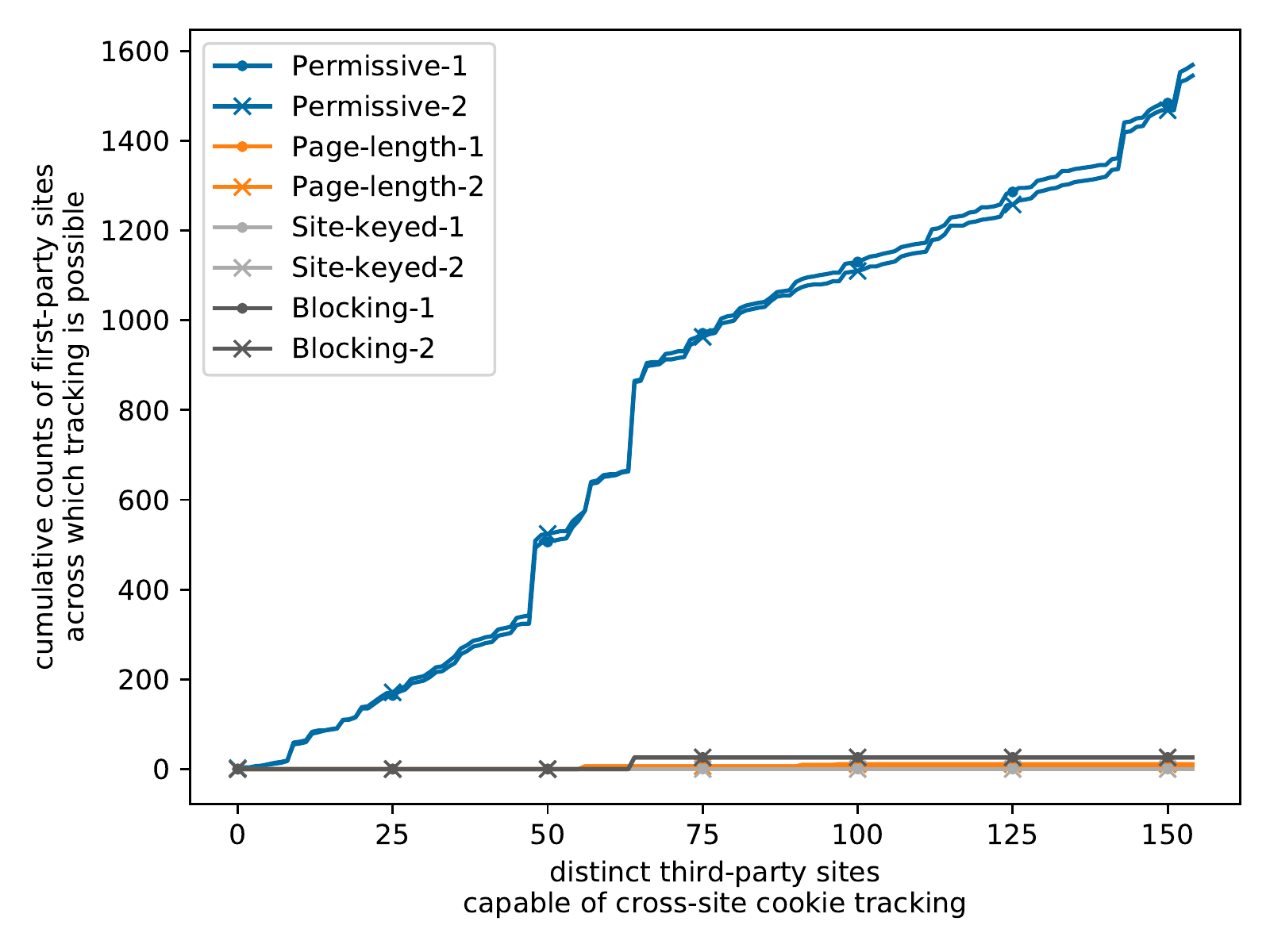}
    \caption{Of our tested policies, all but \efsVanilla{} essentially eliminated stateful cross-site tracking potential.}
    \label{fig:results:privacy:lateral}
\end{figure}

\ToolNameCap{} eliminates stateful cross-site tracking as effectively as does \efsBlock{}.
See Figure~\ref{fig:results:privacy:lateral}.
The cumulative sum curves show the aggregate counts of sites across which \TPs{} could track users under different policies, calculated using the tracking-potential heuristics described in Section~\ref{sec:methodology:evaluation:privacy}.
\efsPrototypeCap{}, \efsSplitKey{}, and \efsBlock{} policies are roughly equal at preventing stateful cross-site tracking.
This result is logical and unsurprising: if \TP{} storage is not available (or is partitioned by \FP{} site, or is strictly ephemeral), it cannot be used to pass identifying state across site boundaries.

\subsection{Privacy: Cross-Time Tracking Potential}
\label{sec:results:privacy:longitudinal}

\begin{figure}[t]
    \centering
    \includegraphics[width=\columnwidth]{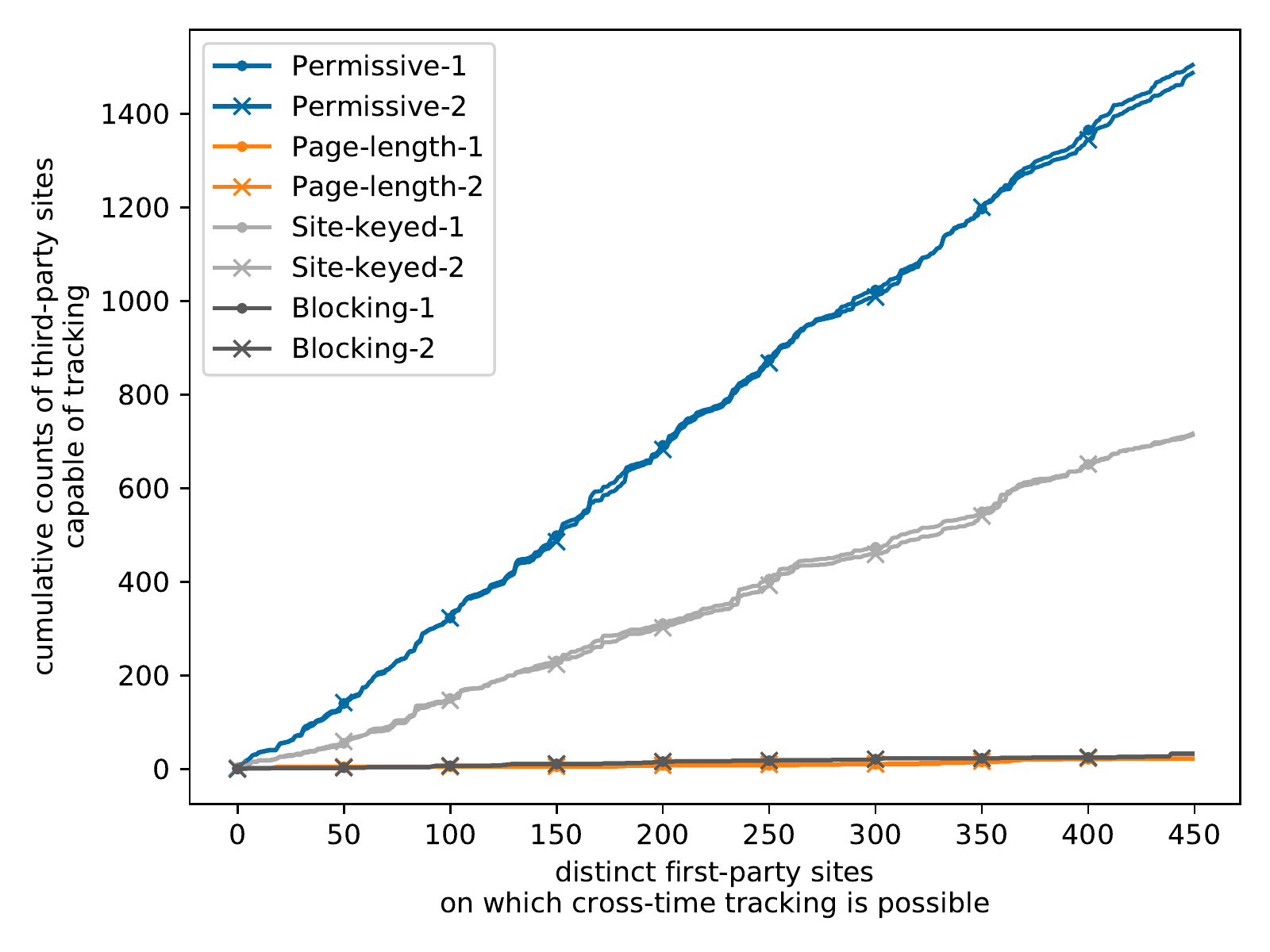}
    \caption{Our \efsPrototype{} policy significantly outperforms both \efsVanilla{} and \efsSplitKey{} policies at reducing cross-time tracking potential.}
    \label{fig:results:privacy:longitudinal}
\end{figure}

\ToolNameCap{} also eliminates stateful cross-time tracking as effectively as does full \TP{} storage blocking, which is a significant improvement over \efsSplitKey{} storage.
See Figure~\ref{fig:results:privacy:longitudinal}.
These curves show the cumulative sums of \TPs{} which could longitudinally track return visitors across the \WebDataSet{} sites, as described in Section~\ref{sec:methodology:evaluation:privacy}.
Unsurprisingly, \efsVanilla{} policy allows the most cross-time tracking; its strong cross-site tracking ability implies cross-time tracking ability.
Persistent \TP{} storage, even if partitioned by \FP{} site context, is still accessible on repeat visits, allowing cross-time tracking.
Thus, \efsPrototype{} and \efsBlock{} policies equally provide stronger cross-time tracking protection than \efsSplitKey{} policy can.

\subsection{Compatibility: Quantitative Assessment}
\label{sec:results:compat:quantitative}

\begin{figure}[t]
    \centering
    \includegraphics[width=\columnwidth]{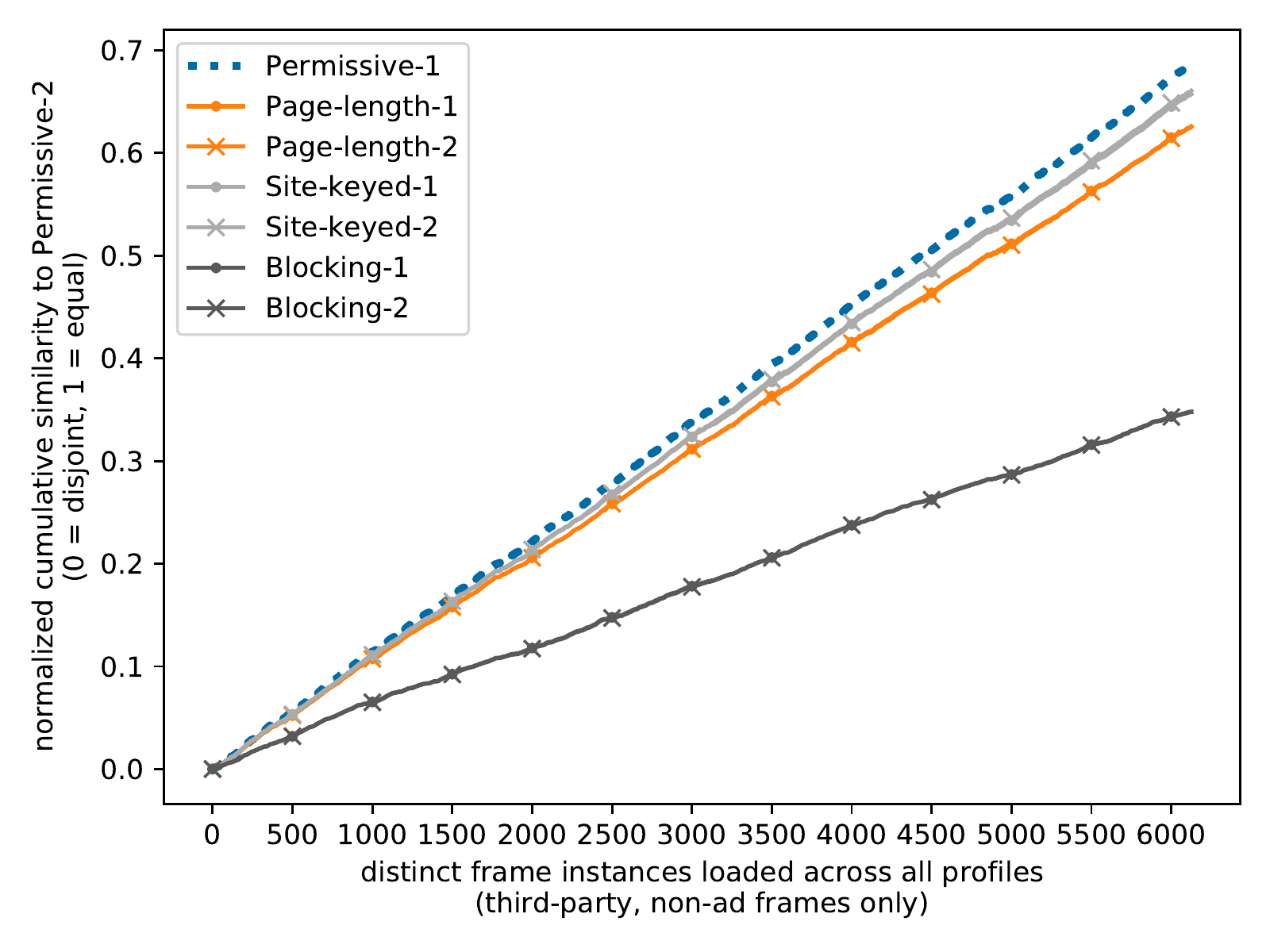}
    \caption{Our \efsPrototype{} policy produces page behaviors within \TP{} frames much closer to the \efsVanilla{} baseline than does the breakage-prone \efsBlock{} policy.}
    \label{fig:results:compat:quantitative-raw}
\end{figure}

\ToolNameCap{} produces page behaviors much closer to the \efsVanilla{} policy baseline than does full \TP{} storage blocking, as shown in Figure~\ref{fig:results:compat:quantitative-raw}.
These curves show cumulative sums of similarity scores between one of our \efsVanilla{} crawl profiles and all other profiles, normalized to show 1.0 as the maximum possible score (perfect similarity on all instances).
The curve showing the similarity scores between the two \efsVanilla{} profiles provides a baseline (\IE{} the best scores observed).
Note the high consistency between all pairs of same-policy profiles.
While even the baseline falls short of perfect similarity, there is a clear signal in the grouping of policies.
The \efsBlock{} policies produced the curves farthest from the baseline, as expected, well isolated from all the other policies.
The non-blocking policies (\efsSplitKey{} and \efsPrototype{}) both produced curves much closer to the baseline than to \efsBlock{}.
The stark separation of curves strongly suggests that the non-blocking policies induce significantly less overall deviation from ``normal'' behavior (and thus less breakage) than does \efsBlock{}.

\subsection{Compatibility: Qualitative Assessment}
\label{sec:results:compat:qualitative}

As described in Section~\ref{sec:methodology:evaluation:compat-qualitative}, we had two graders independently perform manual evaluation on our set of~50 candidate URLs for each of the three profiles: \efsSplitKey{}, \efsPrototype{}, and \efsBlock{} to manually assess each policy's potential for breaking sites. The graders independently evaluated each candidate site for each of the three profiles to find any deviations from our control profile, \efsVanilla{}, the Chrome default. The graders gave each visit a score on a scale of~1 to~3, as detailed in Section~\ref{sec:methodology:evaluation:compat-qualitative}. We conservatively considered deviation from the control visit as a form of breakage, resulting in a score of greater than~1. We summarize the instances of graded breakage for each profile in Table~\ref{table:results:compat:qualitative-breakage}.
Grader notes on several reported breakages for \efsPrototype{} and \efsSplitKey{} suggest that at least some of those deviations involved render process crashes rather than actual content breakage, possibly due to obscure bugs in those prototypes.

\begin{table}[tb]
\centering
\begin{tabular}{l|r|r}
    \hline
    \multicolumn{1}{c|}{\textbf{Profile}} & \multicolumn{1}{c|}{\textbf{\begin{tabular}[c]{@{}c@{}}Pages \\ Broken \end{tabular}}} & \multicolumn{1}{c}{\textbf{\begin{tabular}[c]{@{}c@{}} \% Broken \\ (n=50)\end{tabular}}} \\ \hline
\efsSplitKeyCap{}         & 4                                                                                         & 8\%                                                                                       \\
\efsPrototypeCap{}            & 2                                                                                         & 4\%                                                                                       \\
\efsBlockShort{}            & 5                                                                                         & 10\%                                                                                      
    \end{tabular}
    \caption{Candidate URL breakage as assesses by holistic (whole-page) manual grading}
    \label{table:results:compat:qualitative-breakage}
\end{table}

Considering the 5 breakages observed for the \efsBlock{} profile, the \efsPrototype{} profile either scored similar (2 cases) or improved (3 cases) in terms of raw grader scores.
In contrast to the \efsSplitKey{} profile (4 breakages), the \efsPrototype{} profile again had either scored equal (2 cases) or better (2 cases).
There were \textbf{no cases} where there was a breakage for the \efsPrototype{} profile with worse score compared to either of the \efsBlock{} or the \efsSplitKey{} profiles.
We concluded that the \efsPrototype{} profile performed reliably better than the \efsBlock{} profile, and that it performed as well or better than the \efsSplitKey{} profiles.
The observed rate of breakage for \efsBlock{} (10\%) appears reasonable, and the roughly 2-to-1 advantage of \ToolName{} over \efsBlock{} observed in manual testing parallels a similar advantage in mean cumulative similarity score observed in quantitative analysis (Section~\ref{sec:results:compat:quantitative}).

\section{Discussion}
\label{sec:discussion}


\subsection{Limitations}
\label{sec:discussion:limitations}

The principal design limitation of \ToolName{} is the fact that some useful \TP{} \Web components may simply require persistent, non-partitioned storage.
We suspect that persistent storage for embedded \TP{} content is more a matter of user convenience than essential functionality (\EG{} customizing an embedded video player when the user is logged into the \TP{} site hosting the video).
In any case, \ToolName{} can and should be augmented in production with the \texttt{requestStorageAccess} API to allow the user to opt-in to persistent storage for specific \TPs{}, either universally or on a specific \FP{}.

Our quantitative assessments of tracking and compatibility are subject to the limitations and risks of automated web crawls.
While the scale of our crawl is modest, we believe the \WebDataSet{} provides a realistic sample of popular, mainstream \Web{} content and thus meets our evaluation needs.
Spidering 3 links deep past landing pages likewise provides reasonable sampling of site content without exhausting our time and space budget, as PageGraph can generate large volumes of data per page.
All our crawlers were stateful and non-headless, giving them a fair chance at evading the most trivial forms of bot detection.
More sophisticated bot detection depending on ``human'' interactions with page content should treat all profiles identically (as bots; we performed no interaction simulations).
We thus believe that whatever impact bot detection had on our crawlers, it would have affected all our profiles similarly and not significantly skewed our results.

\subsection{Next Steps}
\label{sec:discussion:future}

\ToolNameCap{} can be further, better evaluated by real users by deploying it first to browsers serving privacy-conscious audiences.
Production implementations will be somewhat more complex than our prototype (to address performance and maintainability concerns) but should require only modest investment by browser vendors.
Production implementations should use the \texttt{requestStorageAccess} API to allow user opt-in to useful \TP{} storage access.
Vendors could then deploy \ToolName{} to privacy-conscious users already blocking \TP{} storage and observe the impact on their site breakage reports.

Ultimately, \ToolName{} can be standardized to provide a near-best-of-both-worlds solution to the problem of persistent \TP{} storage abuse.
Legacy content that assumes \TP{} storage access can be largely accommodated to minimize site breakage, without privacy loss.
Modernized content can use the \texttt{requestStorageAccess} API to bypass \ToolName{} with user consent and gain controlled access to persistent \TP{} storage.
The user wins: content that really needs \TP{} storage access to provide tangible benefit to the user can do so with the user's explicit permission, but the risk of permission denial and user alienation will motivate publishers of content providing less compelling user benefit (\EG{} advertisers and trackers) to make do with less intrusive technologies.

\section{Related Work}
\label{sec:relwork}

\subsubsubsection{Stateful User Tracking}
Storage-based user tracking, usually called ``stateful'' tracking and traditionally involving cookies, has been extensively studied.
Mayer and Mitchell's seminal \TP{} \Web{} tracking study covered both stateful and stateless techniques and introduced the influential FourthParty web measurement framework~\cite{mayer2012fourthparty}.
A contemporary stateful tracking measurement work by Roesner \ETAL{} defined alternate terms ``explicit'' and ``inferred'' for stateful and stateless techniques, respectively, while measuring stateful tracking exclusively~\cite{roesner2012detecting}.
Acar \ETAL{}'s classic, large-scale user tracking measurement study emphasized stateful tracking and hinted at the nascent problem of cookie syncing~\cite{acar2014neverforgets}.
A large-scale evaluation of stateful \TP{} tracking by Li \ETAL{} focused on cookies, the most prevalent form observed, and used machine-learning to identify \TP{} cookie tracking on 46\% of the Alexa 10k~\cite{li2015trackadvisor}.
Engelhardt and Narayanan's extremely large-scale user tracking measurement study included both stateful and stateless techniques, covered the entire Alexa Top Million sites, and introduced the widely used OpenWPM \Web{} measurement framework \cite{englehardt2016openwpm}.
Yang and Yue recently extended classic tracking measurement methodologies to mobile web clients and reported distinctive but analogous groups of tracking domains compared to traditional desktop \Web{} tracking~\cite{yang2020mobilewebtrack}.
Despite the increasing sophistication of \Web{} tracking and countermeasure technologies, Fouad \ETAL{}'s recent exploration of obscure pixel-trackers showed classic \TP{} cookie tracking to still be effective and prevalent in the wild~\cite{fouad2020missed}.
Zimmeck \ETAL{} even found traditional stateful tracking techniques to provide usable building blocks for cross-device tracking via linking together independent tracking sessions from different devices~\cite{zimmeck2017privacy}, a phenomenon conceptually similar to cookie syncing.

\subsubsubsection{Cookie Syncing \& Other State Transfers}
\TPus{} can collude to share stored user tracking identifiers and expand their tracking scope via \textit{cookie syncing}.
Olejnik \ETAL{} performed the first major measurement of cookie syncing in the wild, reporting that up to 27\% of a user's browsing history could be leaked via cookie syncing~\cite{olejnik2014auction}.
Falahrastegar et al. measured distinctive personal identifiers and entities sharing them across the \Web{}, focusing on the groups engaged in sharing and how user behavior affects sharing~\cite{falahrastegar2016tracking}.
Our procedure for selecting potentially identifying cookie flows shares some similarities with their selection of personal identifiers.
Papadopoulos \ETAL{} identified cookie syncing as a major source of hidden costs to users imposed by digital advertising online~\cite{papadopoulos2018cost}.
Subsequent work documented the state of the art in cookie syncing, reinforcing the importance of \TP{} cookies to contemporary tracking~\cite{papadopoulos2019cookiesync}.

Tracking identifiers can be passed across \FP{} domains using means other than stored state.
Stopczynski \ETAL{} provide evidence that modern defenses like Safari ITP are effective but are being actively attacked and evaded, \EG{} via abuse of HTTP redirects passing identifiers in modified URLs~\cite{stopczynski2020redirecttracking}.
For the moment, these attacks appear to constitute efforts to reestablish traditional cookie tracking disrupted by ITP rather than the emergence of a new tracking paradigm.

\subsubsubsection{Browser Fingerprinting}
A major category of \Web{} privacy research for the past decade has involved stateless tracking via fingerprinting.
The Panoptoclick project's seminal report on browser fingerprintability~\cite{eckersley2010panoptoclick} popularized the threat as a potential tracking vector and launched a flurry of related research.
Acar \ETAL{} measured fingerprinting in the wild and found it much more prevalent than commonly estimated at the time~\cite{acar2013fpdetective}.
Olejnik \ETAL{} dissected the infamous and quickly deprecated Battery Status API as a particularly egregious source of fingerprinting entropy~\cite{olejnik2016leaking}.
Laperdrix \ETAL{} identified new fingerprinting vectors from emerging desktop and mobile \Web{} technologies, but also identified potential trends toward reduced fingerprinting threats~\cite{laperdrix2016beauty}.
The current threat status of fingerprinting remains ambiguous: Gomez \ETAL{} reported findings that Panoptoclick-style identification has been largely defeated in practice~\cite{gomez2018hiding}, but Pugliese \ETAL{} later presented counter-arguments from data that such fingerprinting is still an effective threat~\cite{pugliese2020longtermfp}.

\subsubsubsection{Content Blocking}
Published countermeasures against user tracking can be broadly categorized as either blocking tracking-related content (\EG{} ads) before they enter the browser or changing browser implementations to mitigate unwanted effects from such content.
As most ad and tracker blocking currently depends by filter lists, filter list improvements and alternatives are a frequent research topic.
Gugelmann \ETAL{} used large-scale traffic analysis (15k users on a campus network) to train a machine classifier of privacy-invasive tracking services, compared it to popular filter lists, and presented it as a mechanism for updating these lists faster and more effectively than the current crowd-sourced model~\cite{gugelmann2015automated}.
The PageGraph instrumentation system has been used to demonstrate the effectiveness and efficiency of ad blocking via machine-learning trained on page graph data~\cite{iqbal2020adgraph}, to improve filter lists for non-English-speaking populations~\cite{sjosten2020filter}, and to detect filter list evasions in the wild~\cite{chen2021signatures}.
Hu \ETAL{} analyzed the interconnectedness (or ``tangle factor'') of \FP{} sites embedding the same \TP{} tracking content using real-world browsing data from volunteers in order to assess ad blocker effectiveness and to drive automatic partitioning of \FP{} sites into isolated multi-account containers \cite{hu2020websci}.

\subsubsubsection{Browser Policies \& Mechanisms}
\ToolName{} belongs to another category of tracking countermeasure research, which focuses on evaluating and enhancing built-in browser security policies.
Hypothetical discussions of blocking \TP{} storage, and of potential collusion by \TPs{} to work around it, predate the modern era of tracking research~\cite{jackson2006protecting}.
Bauer \ETAL{} demonstrated practical formal browser security using a taint-analysis and data-flow policy enforcement engine build into Chrome 32; the system could be used to enforce classic browser policies (SOP, CSP) or prototype new ones~\cite{bauer2015chromeflow}.
Pan \ETAL{} prototyped an full replacement of the traditional SOP with a hierarchy of nested security principals, each layer able only to increase, not decrease, restrictions on tracking~\cite{pan2015trackingfree}.
Fingerprinting countermeasures that involve injecting randomness into known or suspected entropy sources to disrupt stateless tracking include Privaricator~\cite{nikiforakis2015privaricator} and FPRandom~\cite{laperdrix2017fprandom}.
Yu \ETAL{} described an elegantly generalized approach to tracking prevention at the data flow level using $k$-Anonymity, deployed in the privacy-focused Cliqz browser~\cite{yu2016tracking}.
Our approach to quantifying tracking potential is loosely inspired by this data flow approach to defining privacy.


\section{Conclusion}
\label{sec:conclusion}

Our work addresses the lose-lose dilemma presented to browser developers by \TP{} storage: maintain the status quo and enable mass user tracking, or block \TP{} storage and break a significant amount of the useful web.
Practical experience suggested it was rare for \TP{} content to actually need persistent storage to provide desirable functionality to the user.
We exploited this insight to design \ToolName{}, and the results show that a win-win (or at least a win-nearly-always-win) solution is possible to the old lose-lose dilemma.
We share our contributions with the browser research and development community: the conceptual design of \ToolName{}, a novel solution to the \TP{} state management problem in browsers; our metrics for comparing the privacy and compatibility impact of storage policy changes; our working prototype, made available as open source patches to Chromium, along with our crawl dataset.


{\footnotesize \bibliographystyle{acm}
\bibliography{efs}}

\end{document}